\documentclass[conference]{IEEEtran}
\usepackage{graphicx}
\usepackage{xcolor}
\usepackage{cite}
\usepackage{amsmath,amssymb,amsfonts}
\usepackage{algorithmic}
\usepackage{graphicx,color}
\usepackage{textcomp}
\usepackage{algorithm}
\usepackage{array}
\usepackage[caption=false,font=normalsize,labelfont=sf,textfont=sf]{subfig}
\usepackage{textcomp}
\usepackage{stfloats}
\usepackage{url}
\usepackage{verbatim}
\usepackage{soul}	
\usepackage{multirow}

\ifCLASSINFOpdf
\else
\fi
\begin{document}
\title{Inter-Cell Interference Rejection Based on Ultrawideband Walsh-Domain Wireless Autoencoding}

\author{\IEEEauthorblockN{Rodney Martinez Alonso\\Cel Thys\\Sofie Pollin}
\IEEEauthorblockA{Department of Electrical Engineering\\
KU LEUVEN\\
Leuven, Belgium 3001\\
Email: rodney.martinezalonso@kuleuven.be}
\and
\IEEEauthorblockN{Cedric Dehos}
\IEEEauthorblockA{CEA-Leti\\
Email: cedric.dehos@cea.fr}

\and
\IEEEauthorblockN{Yuneisy Esthela Garcia Guzman}
\IEEEauthorblockA{Silicon Austria\\
	Email: yuneisy.guzman@silicon-austria.com}

}


%Telephone: (800) 555--1212, Fax: (888) 555--1212}


\maketitle

\begin{abstract}
%\boldmath
This paper proposes a novel technique for rejecting partial-in-band inter-cell interference (ICI) in ultrawideband communication systems. We present the design of an end-to-end wireless autoencoder architecture that jointly optimizes the transmitter and receiver encoding/decoding in the Walsh domain to mitigate interference from coexisting narrower-band 5G base stations. By exploiting the orthogonality and self-inverse properties of Walsh functions, the system distributes and learns to encode bit-words across parallel Walsh branches. Through analytical modeling and simulation, we characterize how 5G CP-OFDM interference maps into the Walsh domain and identify optimal ratios of transmission frequencies and sampling rate where the end-to-end autoencoder achieves the highest rejection. Experimental results show that the proposed autoencoder achieves up to 12 dB of ICI rejection while maintaining a low block error rate (BLER) for the same baseline channel noise, i.e., baseline Signal-to-Noise-Ratio (SNR) without the interference.
\end{abstract}


% \fi
\IEEEpeerreviewmaketitle

\section{Introduction}
Each new application or use case enabled by wireless technologies has quickly run into system bandwidth as the critical bottleneck. Nowadays, when the spectral efficiency of the latest developments in wireless technologies is close to 1~dB from the Shannon limit, and some techniques report a fraction of decibel difference~\cite{CloseToShannon}, there is not too much margin of improvement on  per-channel spectral efficiency. Optimization techniques to offload traffic, such as multilayered edge caching~\cite{EdgeCaching}, and multi-service offloading in layer division multiplexing~\cite{LDMrodney} have not been able to slow down the growth of bandwidth demand.

Spectrum utilization efficiency has become fundamental to address the never-ending bandwidth demand. Paradoxically, several spectrum surveys show that even on the \textit{scarce} sub-6-GHz spectrum, the effective use of it at a given location and instant of time can be as low as 5\% on average~\cite{SpectrumSurvey1}. Even in the case of sub-1-GHz it can be as low as 11\%~\cite{SpectrumSurveyRMA}. In this sense, although technologies like carrier aggregation allow combining non-contiguous frequency bands, this leads to inefficiencies in spectral use due to band guards, increased complexity of RF transceivers, and synchronization overhead. Aggregating multiple channels across different bands generally requires multiple RF chains, increasing also the power consumption. In addition, due to technological transitions and spectrum re-farming, this also has an indirect impact on shortening the transceiver hardware life cycle. 

The main challenges for multi-band carrier aggregation are precisely the non-uniform characteristics of the aggregated bands and strict synchronization requirements. For instance, authors in~\cite{CA1} show that aggregating bands with different numerologies (i.e., subcarrier spacing, symbol duration, and cyclic prefix length) leads to synchronization issues and capacity losses compared to the equivalent continuous-band transmission. Addressing multi-band physical layer reconfiguration with software-defined radios (SDRs) has been proposed to handle the difference in radio-propagation characteristics for multi-band carrier aggregation. In this case, a flexible SDR solution could adapt the transmission bandwidth, modulation, and encoding schemes by modifying the virtualization of the digital signal processing. However, this flexibility comes at a cost in terms of increased power consumption~\cite{CA2}. 

In the case of the RF chain for multi-band signal processing, a critical gap in the state-of-the-art is the lack of an efficient multi-band and adaptable Low Noise Amplifier~(LNA) solution. On one hand, reconfigurable LNAs are not energy efficient, while multi-band LNAs require impedance transformers for input matching across different frequency bands, which increases the chip size and also power consumption~\cite{CA2}.

A more spectral efficient approach would be to process the bandwidth in a continuous manner. However, ultrawideband transceivers (e.g., bandwidth $\mathit{B_w \geq 1 GHz}$) have significantly lower energy efficiency. The primary challenge and bottleneck lie in the data converters, since their power consumption, noise, and cost rise with sampling frequency. Indeed, a large-scale survey on Analog-to-Digital-Converters (ADC) from a systematic literature review of the state-of-the-art, comprising more than 700 ADC designs, shows that the growth rate of power consumption per conversion step as a function of sampling frequency (i.e., Walden figure of merit) increases from approximately 250~MHz onward. To improve the figure-of-merit of the data conversion, authors in~\cite{Walsh1} have proposed an interleaved converter based on Walsh transformation. This converter operates in frequency domain but with lower complexity than Fourier-based architectures, replacing sinusoidal basis functions with Walsh-Hadamard binary wavelets (i.e., $\mathit{+1}$ or $\mathit{-1}$ pulses). Compared to time-interleaving converters, the Walsh converter architecture has the advantage of removing the interleaving multiplexer, which is a major contributor to the dynamic power consumption and a source of impairments. The Walsh transceiver with 64 branches presented by the authors in~\cite{CedricWalsh} achieved a sampling rate of 10~GSPS with a power consumption of only 48~mW, which is at least half compared to the state-of-the-art in terms of absolute power. This hardware architecture could even be implemented transparently regarding the primary signal modulation (e.g., OFDM).

Increasing instantaneous bandwidth to achieve higher throughput faces technical challenges not only in terms of transceiver energy efficiency but also in terms of spectrum allocation. Despite the fact that several fragments of the spectrum are not occupied most of the time in all locations, there is an sparse use of it across legacy services. Therefore, the focus of our research in this paper is how a Walsh-based ultrawideband communication system could coexist with legacy technologies, considering a more aggressive reuse of the spectrum. The novelty of our paper lies in the development of a technique to reject the interference from partial-in-band coexisting 5G base stations in a Walsh-based ultrawideband link. Specifically, we exploit the property of the Walsh transformation to asymmetrically distribute the interference signal in the branch processing chains. To that purpose, our algorithm exploits the data parallelization of an adapted version of the ultrawideband transceiver developed in~\cite{CedricWalsh} with an AI-based End-to-End wireless autoencoder. The End-to-End wireless autoencoder allows encoding message bits into the Walsh domain in an interference-aware manner. This means that the system learns an encoding function and equalization that optimally exploit the Walsh branches that are less impacted by the interference. 

The outline of this paper continues as follows: in Section~\ref{sec:TheMethod} we present the general rationale of our research, followed by the details of the Walsh-domain End-to-End autoencoder in subsection~\ref{subsec:E2EWalsh} and the scenario and interference modeling in subsection~\ref{subsec:5GModel}. The results and numerical analysis are presented in Section~\ref{sec:Results}, which include a characterization of the interference pattern distribution in the Walsh domain, an assessment of how much our system improves the Block Error Rate (BLER) for a given Inter-Cell Interference (ICI) margin, and the ICI rejection as a function of the interference frequency.

\section{Method}
\label{sec:TheMethod}
\begin{figure}[h]
	\centering
	\includegraphics[width=0.95\linewidth]{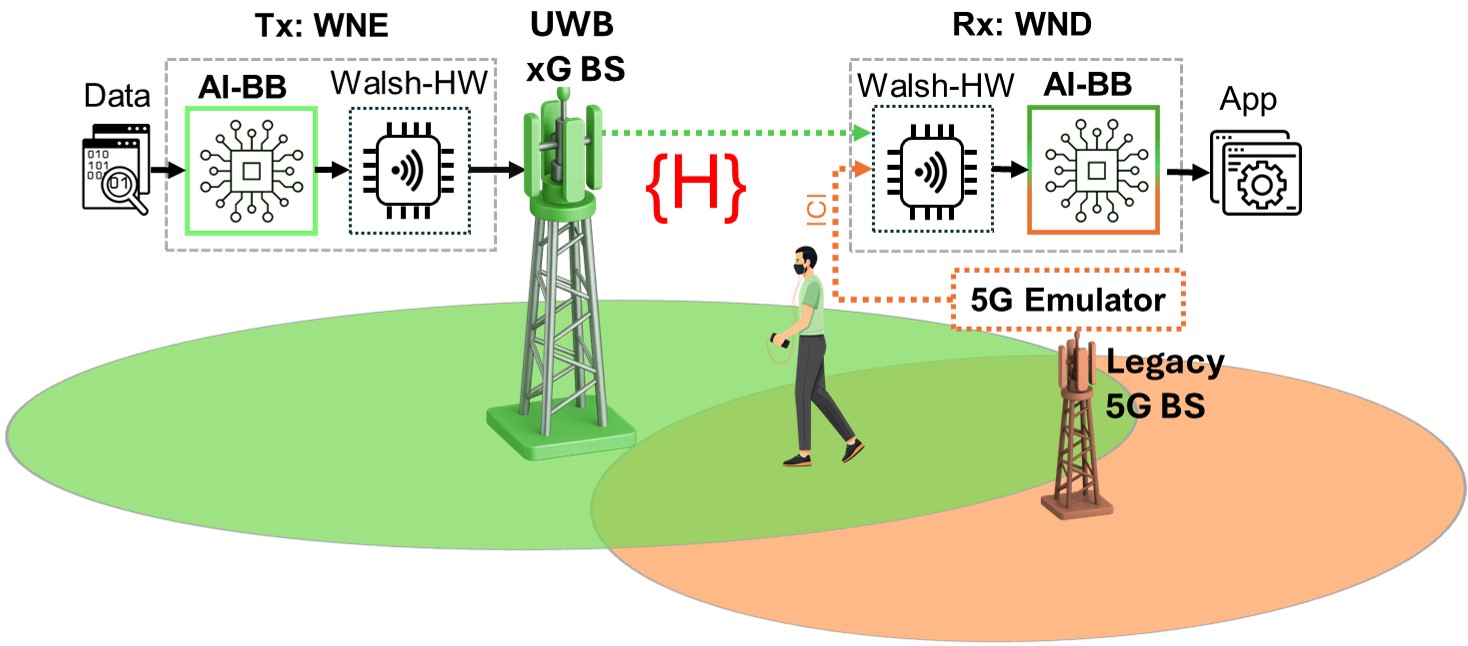}
	\caption[Research Method Rationale]{High-Level Diagram of Research Method Rationale}
	\label{fig:research-method-rationale}
\end{figure}

The methodology of our research aims to assess the capability of an end-to-end wireless interface to encode the data in Walsh domain in a way that mitigates the effect of inter-cell interference from legacy 5G services reusing part of the spectrum. The scenario represented in Figure~\ref{fig:research-method-rationale} shows an UWB next generation (xG) Base Station (UWB xG BS) supporting slightly less than 2.5 GHz instantaneous bandwidth. At the physical wireless interface, this bandwidth is achieved by a Walsh-based hardware transceiver that allows easing the sample rate requirements by parallelizing the signal processing chain. The transmitted signal is further distorted by the transmission channel. We consider an Additive White Gaussian Noise channel (AWGN) to model how the signal is affected by random baseline noise ($\mathit{\{H\}}$ in Figure~\ref{fig:research-method-rationale}). On the receiver side, an equivalent transceiver recovers the Walsh domain signal and decodes the Walsh-encoded symbols into estimated message bits. In the receiver, the UWB signal is also distorted by different levels of Inter Cell Interference~(ICI) from a legacy 5G BS that shares part of the UWB spectrum.

\subsection{End-to-End Wireless Autoencoder in Walsh domain}
\label{subsec:E2EWalsh}
\begin{figure}[h]
	\centering
	\includegraphics[width=0.95\linewidth]{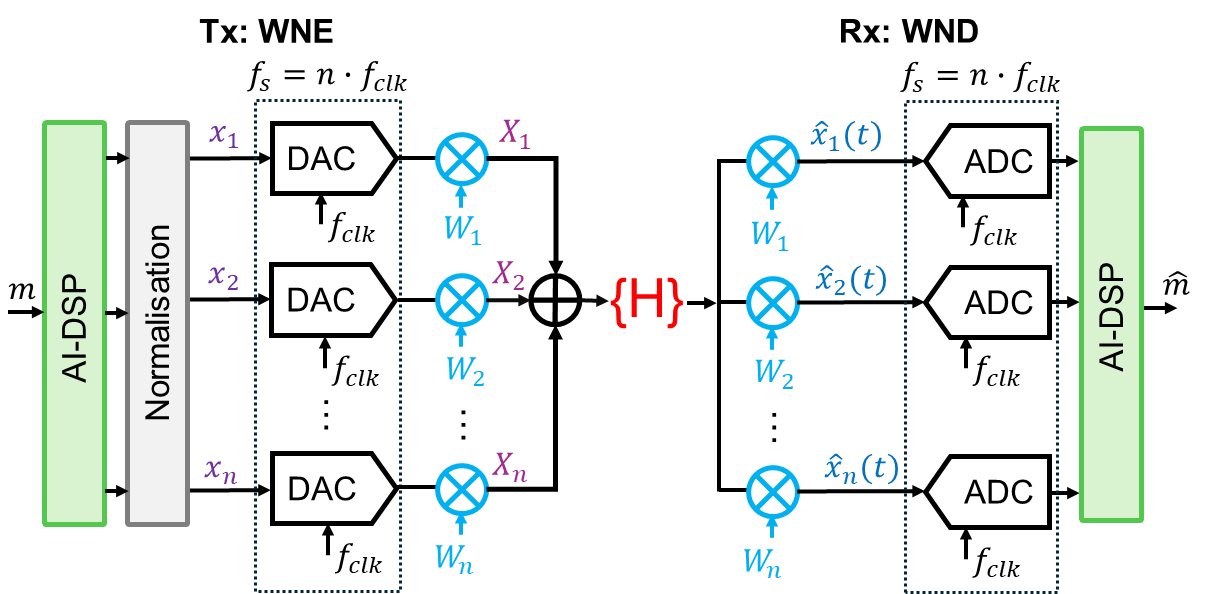}
	\caption[Walsh E2E Transceiver]{Diagram of the transceivers in the End-to-End wireless autoencoder}
	\label{fig:walshe2etransceiver}
\end{figure}

Figure~\ref{fig:walshe2etransceiver} shows a diagram of the transceivers at the base band level in the architecture of the End-to-End wireless auto-encoder (from transmitter to receiver). For clarity, some analog components like integrator, sample and hold, and the RF components are omitted.

The communication link between the transmitter and the receiver works as an end-to-end wireless autoencoder of the physical layer~\cite{CEL_PREPRINT,oshea_hoydis_introduction_2017}. In the end-to-end autoencoder, a Walsh Neural Encoder (WNE) and Walsh Neural Decoder (WND) are jointly trained to map and encode the message bits ($\mathit{m}$) into Walsh symbols (and the reverse operation) in a way that minimizes the BLER caused by the distortions in the channel $\mathit{H}$. Rather than an in-quadrature-based mapping of the constellation, in the design of our proposed WNE each possible message $\mathit{m}$ of $\mathit{k}$ bits is mapped onto $\mathit{N}$ real symbols, transmitted over $\mathit{N}$ real channel uses. As the projection into the Walsh domain is based on real-valued symbols, this results in a coding rate of $\mathit{R=k/N}$ message bits per channel use. In our design, we consider $\mathit{k=4}$ and $\mathit{n=32}$. Subsequently, 2-channel uses are mapped into a symbol, which results in 2-channel uses per symbol.  

\subsubsection{Walsh Neural Encoder layers}

The unsupervised AI model in the transmitter comprises an input and a fully connected layer of size $\mathit{M=2^k=16}$, followed by another fully connected layer of size $\mathit{N}$, for mapping the symbols into the real channel uses. After each fully connected layer, a ReLu activation function is used.

\subsubsection{Walsh domain projection in the Transmitter}

The orthogonality of the channel realizations is based on the Walsh transformation. The Walsh transformation will represent the encoded signal message as a weighted sum of orthogonal, binary-valued basis functions. Unlike the Fourier transform, which uses continuous sinusoidal basis functions, the Walsh transform uses discrete pulses with amplitude $\mathit{+1}$ or $\mathit{-1}$. A key advantage is the simplicity of the Walsh basis, allowing low complexity and efficient representation in the analog domain. The Walsh transformation is critical in our design for reducing the required sampling rate of the data converters while retaining orthogonal transmissions. The Walsh transceiver enables an instantaneous bandwidth of 2.5 GHz by means of 32 parallelized processing branches. This allows reducing the per-branch sampling frequency by $\mathit{1/32}$. This means for a sampling rate requirement of 5~Giga-Samples-Per-Second (GSPS), the per-branch sampling rate is reduced to only 156.25~MSPS.

The Walsh transformation of a finite signal is defined as:

\begin{equation}
	X_{i} = \int_0^T x(t) \cdot W_{i}(t) dt \quad i=1,\dots,N
	\label{Walsh_Transformation}
\end{equation}
where $\mathit{X_{i}}$ represents the weighted contribution of the corresponding Walsh function $\mathit{W_{i}(t)}$
to form the signal $\mathit{x(t)}$. The Walsh functions $\mathit{W_{i}(t)}$ are a complete set of orthogonal basis functions from $[0,T]$ consisting of binary pulses\cite{WalshMaxandre,thys_walsh-domain_2024,CEL_PREPRINT}. Therefore, the vector $\mathit{\mathbf{X}_N}$ is the mathematical projection of the time-domain signal into Walsh-domain. The Walsh transformation  can be discretized as a vector by matrix multiplication:

\begin{equation}
	\mathit{\mathbf{X}_{N}} = \mathit{\mathbf{x}_N} \cdot W_N
	\label{Walsh_Transformation_Digital}
\end{equation}
where the size of the vector of time domain samples $\mathit{x}$ is limited at each time interval to $\mathit{N}$. A fundamental property of the Walsh transform is that, up to a scalar factor, it is an involution, i.e., it is self-inverting. In other words, the transform matrix $W_N$ is a square $\mathit{N}$-by-$\mathit{N}$ symmetric matrix that is its own inverse. The self-inverse property means that multiplying by the same Walsh matrix will encode or decode the signal into the Walsh domain and vice versa.

\subsubsection{Transmission Channel}
The transmission channel is assumed an AWGN channel with noise variance $N_0$:
\begin{equation}
    \mathit{\mathbf{\hat{X}}_N} = \mathit{\mathbf{X}_N} + \mathit{\mathbf{Z}_N},
\end{equation}
where $\mathit{\mathbf{Z}_N}$ is a vector of independent identically distributed samples of thermal noise: $Z_i \sim \mathcal{N}(0,N_0)$. We ensure a consistent Signal-to-Noise-ratio (SNR) or energy-per-bit to noise variance ratio ($E_b/N_0$) by using a normalisation layer at the transmitter~\cite{oshea_hoydis_introduction_2017,CEL_PREPRINT}. 

\subsubsection{Receiver signal recovery}

In the receiver, rather than a real-based converter, we exploit the Walsh transform self-inverse property in order to decode the single input channel by a parallelized signal processing chain and reduce the sampling requirement of the analog to digital converter in a similar way as on the transmitter side. In this case, the data acquisition must be done in parallel in the analog domain, as the signal is received as a single input channel. 

The analog transceiver will implement part of the \textit{inverse} Walsh transformation in the analog domain by multiplying the received signal by each of the Walsh basis functions in an array of analog multipliers (or switches). Formally:

\begin{equation}
\begin{matrix}
	\hat{x}_i(t) = \hat{X}_{i} \cdot W_{i}(t) &   i=1,\dots,N\\
\end{matrix}
	\label{Inverse_Walsh_analogue}
\end{equation}
This will generate $\mathit{N=32}$ "slower-down" time-domain streams ($\hat{x}_i(t)$) for further reconstructing the time domain signal ($\hat{x}(t)$) in the digital domain:

\begin{equation}
	\hat{x}(t) = \sum_{i=1}^{N} \hat{X}_{i} \cdot W_{i}(t)
	\label{Inverse_Walsh}
\end{equation}

\subsubsection{Walsh Neural Decoder layers}

In the receiver, the signal reconstruction and received symbol estimation is performed by an unsupervised machine learning model similar to the transmitter side. In the receiver side, the first layer of the machine learning after the analogue to digital converters is a fully connected layer with $\mathit{N}$ inputs and $\mathit{M}$ outputs, followed by a ReLU activation. The second layer is a fully connected layer with $\mathit{M}$ inputs and $\mathit{M}$ outputs. The M-dimensional output vector is then transformed by a softmax layer into normalized probabilities. This probabilistic representation allows to calculate the cross-entropy loss for end-to-end training of the autoencoder via stochastic gradient descent. Here, we implicitly assume the architecture includes a feedback loop to send the gradients from the decoder to the encoder (during training). During inference, each transmitted message is recovered in the receiver by selecting the message bits $\mathit{\hat{m}}$ with the highest estimated probability from the softmax function.

\subsection{5G interference modeling}
\label{subsec:5GModel}
For the inter-cell interference scenario depicted in Figure~\ref{fig:research-method-rationale}, we emulate a 5G OFDM waveform for each interference sub-band from the 5G NR standard FR2 (ETSI TS 138 104). The 5G signal frequency is processed in our model at the base band level to abstract from the actual spectrum assignment. For generating the 5G signal and accurately fingerprinting its Walsh patterns, we use a Matlab emulator of the Walsh receiver hardware designed in~\cite{CedricWalsh}. This emulator allows extracting the mean absolute variation of different interference signals in the Walsh domain. In this paper, we do not evaluate the effect of the UWB service over the 5G legacy deployment, as we assume the UWB signal has a lower radiated power due to power amplifier constraints. We assume the SNR of the 5G service is not significantly impacted by adjacent UWB cells.

To model the OFDM partial-in-band interference, we distribute random Gaussian noise~\cite{OFDM_gaussian_approximation} into the Walsh branches following a mask filter generated from the emulator output (i.e., additive Gaussian noise shaping). The vector representing the received signal affected by the interference $\mathit{\hat{X}'_N}$ is then calculated as:

\begin{align}
		\hat{X}'_i &= \hat{X}_i + I_i \odot \eta, \quad i=1,\dots,n \\
        \eta &\sim \mathcal{N}(0,1),
	\label{Interference_added}
\end{align}
where the vector $\mathit{\mathbf{I}_N}$ is a mask representing how the ICI power is distributed in the Walsh domain and $\mathit{\eta}$ models the interference as additive Gaussian noise. $\mathit{I_N}$ also scales the ICI power above the baseline channel noise level (from $\mathit{H}$). The specific parameters of the interference scenario and 5G waveform parameters are listed in the following table.

\begin{table}[h]
	\centering
	\caption{Scenario and 5G ICI emulation}
	\begin{tabular}{llc}
		\hline
		\textbf{} & \textbf{Parameter} & \textbf{Typical Value} \\
		\hline
		\multirow{3}{*}{Scenario} 
		& xG Bandwidth  & 2.5 GHz \\
		& 5G Interference Bandwidth & 50 MHz \\
		& \multirow{2}{*}{ICI Base Band Frequency $(F_c)$} & 1250--1650 MHz \\ 
		&                           & 2075--2475 MHz \\
		& ICI Level & 1--12 dB \\
		\hline
		\multirow{3}{*}{Walsh Rx}
		& Number of Branches $(n)$ & 32 \\
		& Noise Floor & -140 dBc/Hz \\
		& ADC bits & 8 \\
		& ADC per branch clock $(F_{clk})$ & 156.25 MHz \\
		& System Sampling $()F_{s})$ & 5 GHz \\
		\hline
		\multirow{6}{*}{5G-FR2}
		& Modulation & CP-OFDM \\
		& Mapping & 64-QAM \\
		& Used Bandwidth & 47.52 MHz \\
		& Nyquist rate & 61.44 MHz \\
		& Symbol duration & 20.83 $\mu$s \\
		\hline
	\end{tabular}
\end{table} 

The xG Bandwidth specifies the total bandwidth where channels coexist, i.e., the instantaneous wanted channel bandwidth. The 5G Interference bandwidth is the bandwidth of the nearby 5G interference cell that overlaps with part of the wanted channel. We modeled a range of interference above the baseline channel noise level of 1 to 12 dB, and considered a central frequency range for the interference between 1250 and 1650 MHz and between 2075 and 2475~MHz with 100~MHz steps inside the range. Notice that this frequency range is around the $\mathit{1^{st}}$ and $\mathit{2^{nd}}$ sub-multiples of the UWB system sampling frequency $\mathit{f_s}$. For the Walsh receiver parameters that affect our system modeling, we consider a real hardware implementation as reported in~\cite{CedricWalsh}. For the 5G-FR2 waveform, the modulation scheme is CP-OFDM with 1024 subcarriers, and the mapping uses a 64-QAM constellation. The total used channel bandwidth corresponds to 792 active subcarriers, while the Nyquist rate considers all 1024 subcarriers.

\section{Results}
\label{sec:Results}
\subsection{Characterisation of Interference Pattern in Walsh Domain}
\begin{figure}[h]
	\centering
	\includegraphics[width=0.95\linewidth]{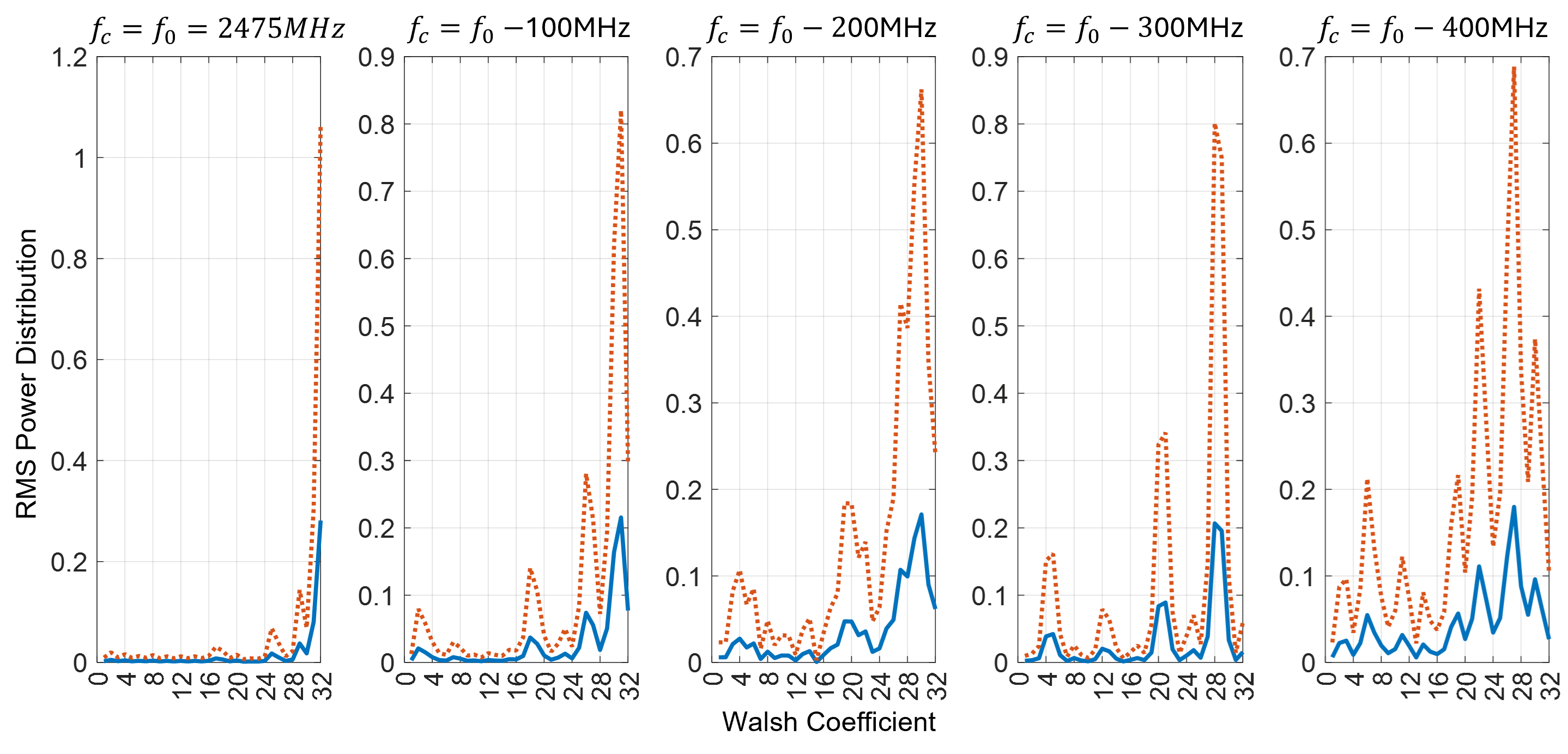}
	\caption[Interference Pattern N=1]{Interference pattern in Walsh domain from 2475~MHz downwards}
	\label{fig:interferencepatternn1}
\end{figure}

Figure~\ref{fig:interferencepatternn1} illustrates the peak (dotted orange line) and mean power distribution (blue continuous line) in the Walsh domain for 5G signals with a 50~MHz bandwidth, centered at the frequency closest to the Nyquist frequency $\mathit{F_{Nyq}= f_s/2^1}$. For comparison, the same signal is projected at different center frequencies, each separated by 100 MHz.

When the interference center frequency is near the Nyquist limit ($\mathit{F_{Nyq}= f_s/2^1}$), most of its power is concentrated in only a few Walsh coefficients. As the center frequency shifts further away from this point, the power becomes gradually spread over a larger number of coefficients. For variation of only -100 MHz (4\% compared to the vicinity of $\mathit{F_{Nyq}}$), the mean absolute deviation from the peak power decreases by a factor of 1.4. A lower mean absolute deviation from the Walsh branch with higher power means that the energy is more spread across the different Walsh coefficients. At -400 MHz, the mean absolute deviation is already decreased by a factor of 2.

\begin{figure}[h]
	\centering
	\includegraphics[width=0.95\linewidth]{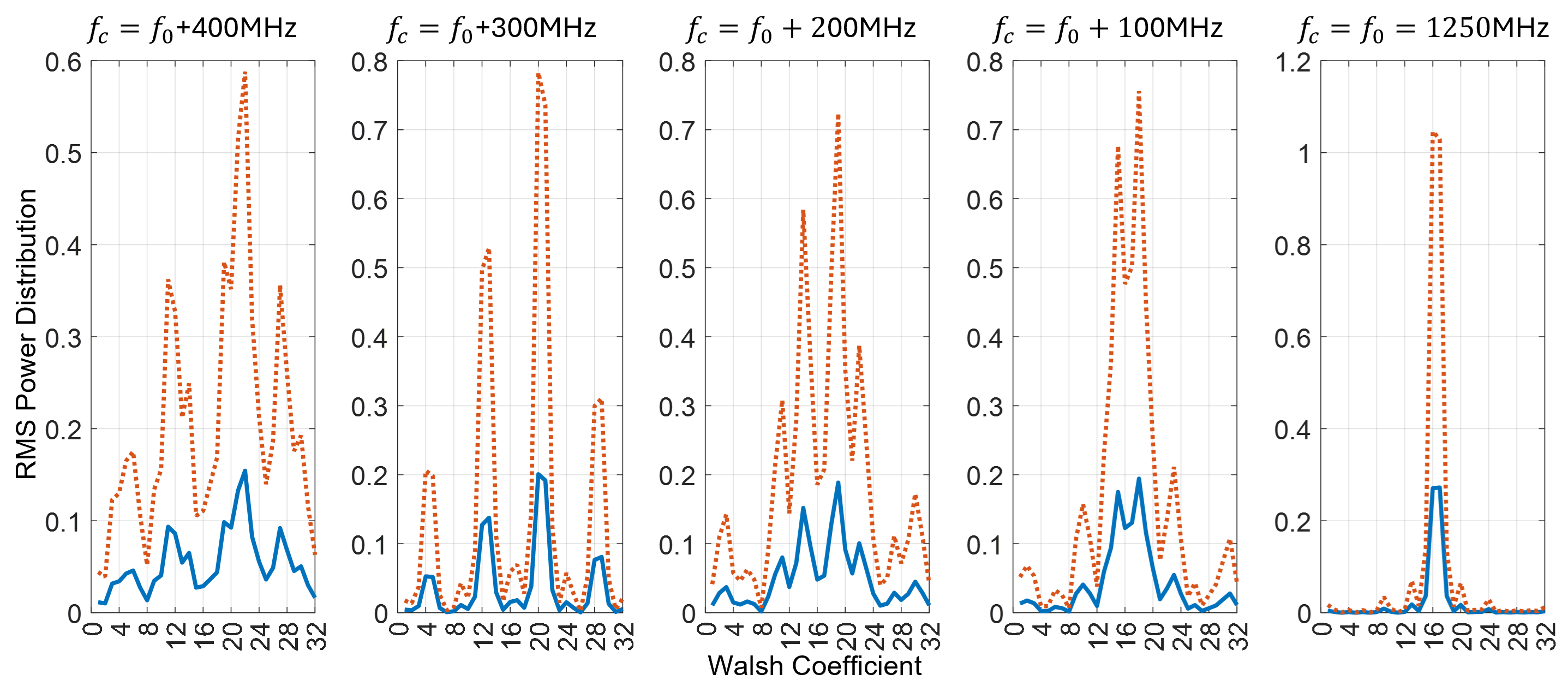}
	\caption[Interference Pattern N=2]{Interference pattern in Walsh domain downwards from 1250 MHz}
	\label{fig:interferencepatternn2}
\end{figure}

As the center frequency gets closer to the next submultiple of the sampling frequency (i.e., $\mathit{F_{c}=f_s/2^2}$), the power starts concentrating again on a few Walsh coefficients. Figure~\ref{fig:interferencepatternn2} shows the peak and mean power distribution in the Walsh domain for 5G signals of 50 MHz as the center frequency approaches $\mathit{F_{c}=f_s/2^2=1250 MHz}$. At $\mathit{F_{c}=f_s/2^2}$, the power concentrates into a couple of Walsh coefficients again, with the mean absolute deviation from peak only 3.8\% lower than for the vicinity of $\mathit{F_{Nyq}}$. At +100 MHz from $\mathit{F_{c}=f_s/2^2}$ (i.e., 1350 MHz in base band), the mean absolute deviation compared to the Walsh vector corresponding to the vicinity of $\mathit{F_{Nyq}}$ is lower by a factor of 1.6. However, this is still better than at 2075 MHz (near $\mathit{F_{Nyq}-400MHz}$) by 20\%.

\begin{figure}[h]
	\centering
	\includegraphics[width=0.85\linewidth]{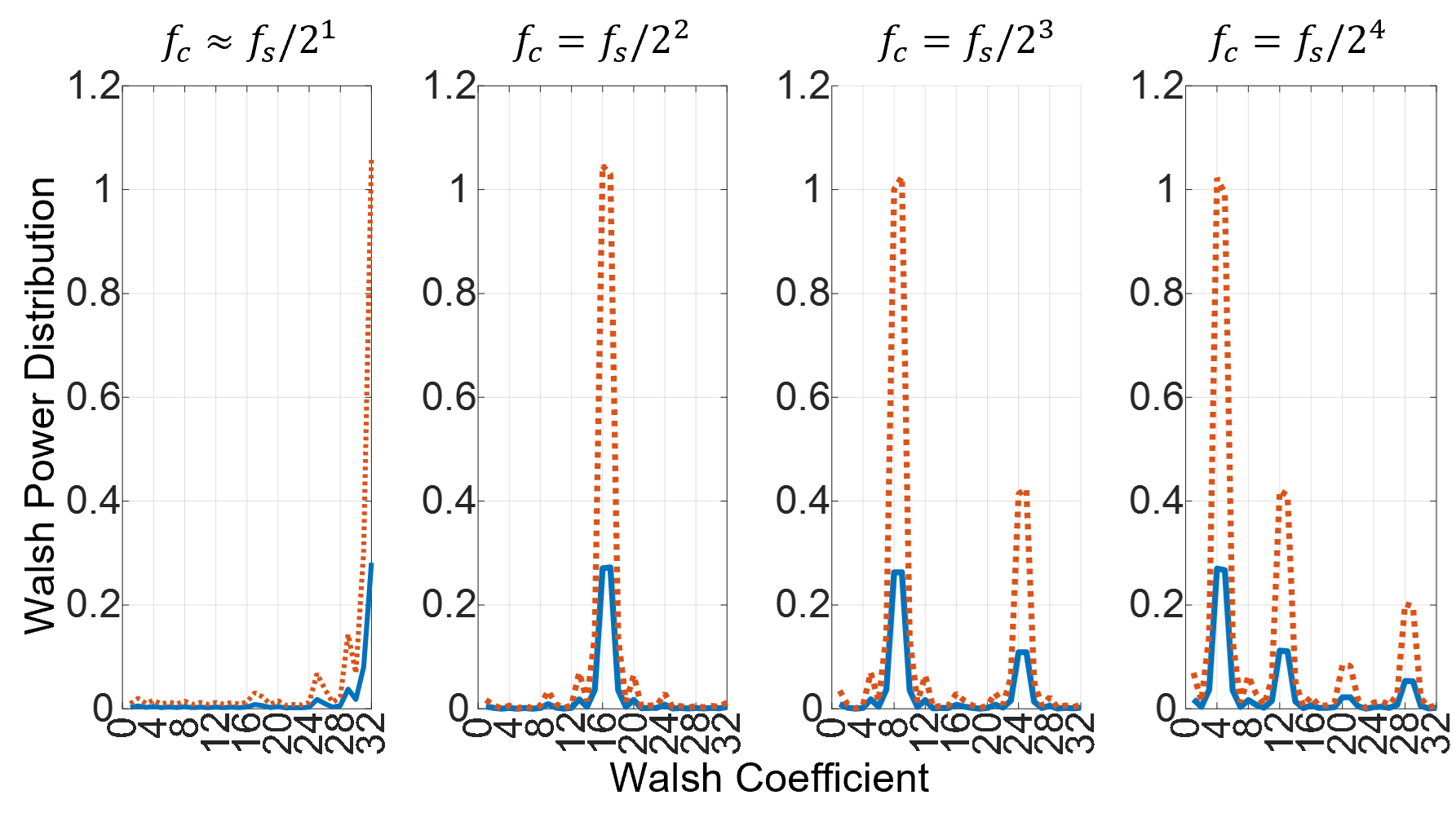}
	\caption[Interference Pattern all submultiples]{Interference pattern in Walsh domain for different submultiples of the sampling frequency}
	\label{fig:interferencepatternnall}
\end{figure}

The interference pattern analysis clearly shows that for interference centered on sub-multiples of the sampling frequency and near them, most of the power concentrates on a few Walsh coefficients. However, the mean absolute deviation decreases as the dividing factor of the sub-multiple $\mathit{2^N}$ increases, and therefore, the power spreads more across Walsh branches. As shown in Figure~\ref{fig:interferencepatternnall}, the mean absolute deviation from peak is 1.0002, 0.9619, 0.9113, and 0.8924 for $\mathit{N={1,2,3,4}}$, respectively. The average gradient of decay as a function of N is -3.7\%.

\subsection{BLER Improvement for a given ICI margin}
\begin{figure}[h]
	\centering
	\includegraphics[width=0.95\linewidth]{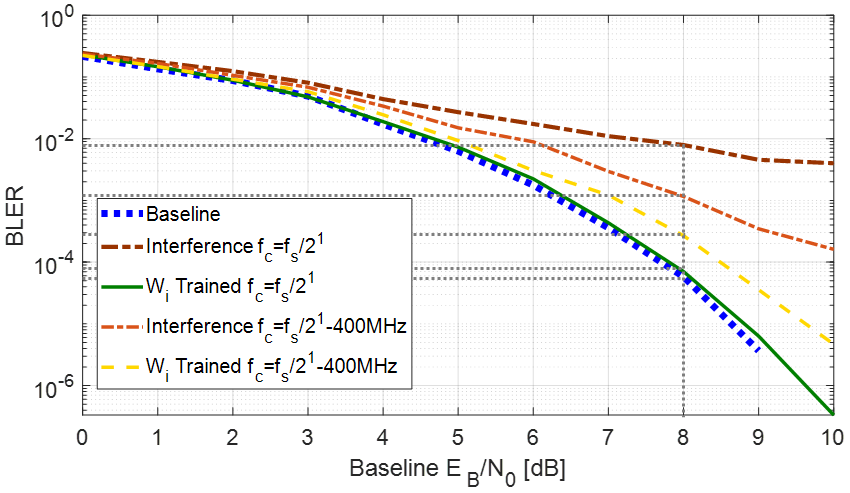}
	\caption[BLER improvement for N=1]{BLER improvement by the End-to-End autoeconder for an interfering 5G signal in the vicinity of $\mathit{f_s/2^1}$, with an ICI of 6~dB}
	\label{fig:bler-vs-ebn0-n1}
\end{figure}

\begin{figure}[h]
	\centering
	\includegraphics[width=0.95\linewidth]{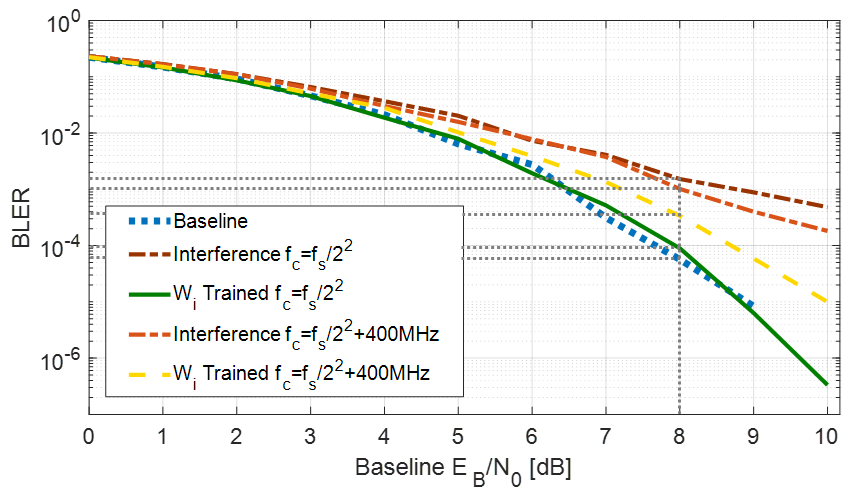}
	\caption[BLER improvement for N=1]{BLER improvement by the End-to-End autoencoder for an interfering 5G signal in the vicinity of $\mathit{f_s/2^2}$, with an ICI of 6~dB}
	\label{fig:bler-vs-ebn0-n2}
\end{figure}

Figure~\ref{fig:bler-vs-ebn0-n1} and Figure~\ref{fig:bler-vs-ebn0-n2} shows the BLER as function of the $\mathit{E_B/N_0}$, for the vicinity of $\mathit{Fs/2^1}$ and $\mathit{Fs/2^2}$, respectively and considering an ICI margin of 6 dB. In these Figures, we isolate the effect of the noise channel from the interference to quantify the added impact of ICI on the system performance.

In Figure~\ref{fig:bler-vs-ebn0-n1} the BLER is in the range between $\mathit{4 \cdot 10^{-5}}$ and $\mathit{6 \cdot 10^{-5}}$ at $\mathit{E_B/N_0=8 dB}$ for the channel noise baseline (without interference). The noise baseline range variability is due to the model error for the different realizations. The added intercell interference of 6~dB causes the BLER to increase to approximately  $\mathit{1.2 \cdot 10^{-3}}$ for $\mathit{f_c\approx f_s/2^1-400~MHz}$ (i.e., equivalent to 2075 MHz in baseband) and up to $\mathit{7.9 \cdot 10^{-3}}$ for $\mathit{f_c\approx f_s/2^1}$. Although the BLER performance is worse near $\mathit{f_c\approx f_s/2^1}$, the end to end-to-end autoencoder has a higher capability to reject and mitigate the interference, due to the fact that the energy concentrates only on a few Walsh branches. Indeed, for $\mathit{f_c\approx f_s/2^1}$ (continuous green line in Figure \ref{fig:bler-vs-ebn0-n1}), the BLER is almost the same as the baseline model (i.e., the difference is within the model error standard deviation). When the interference is at 400~MHz from  $\mathit{f_c\approx f_s/2^1}$, the end-to-end autoencoder is not that effective, but still manages to reduce the BLER by a factor of 4.3 (cf. yellow dashed line in Figure \ref{fig:bler-vs-ebn0-n1}).

Figure~\ref{fig:bler-vs-ebn0-n2} shows the autoencoder capability to reject the interference in the next sub-multiple (i.e., $\mathit{f_c=f_s/2^2}$). In the case of the next sub-multiple of $\mathit{f_s}$, the end-to-end autoencoder is as effective as for the first sub-multiple (i.e., $\mathit{f_s/2^1}$) with the model trained to cancel the interference in the Walsh domain, achieving virtually the same performance as the noise baseline model (without the 5G interference). At +400 MHz from $\mathit{f_c=f_s/2^2}$, the model trained to reject the interference also achieved similar performance than the equivalent model from Figure~\ref{fig:bler-vs-ebn0-n1} (cf. dashed yellow line). An interesting finding is that as the energy is not so concentrated in a few Walsh branches as in the previous figure, the impact on the models not trained to reject the interference is slightly lower (cf. orange and red lines in Figure~\ref{fig:bler-vs-ebn0-n1} and Figure~\ref{fig:bler-vs-ebn0-n2}).

\subsection{BLER vs Interference Frequency}

\begin{figure}[h]
 	\centering
 	\includegraphics[width=0.95\linewidth]{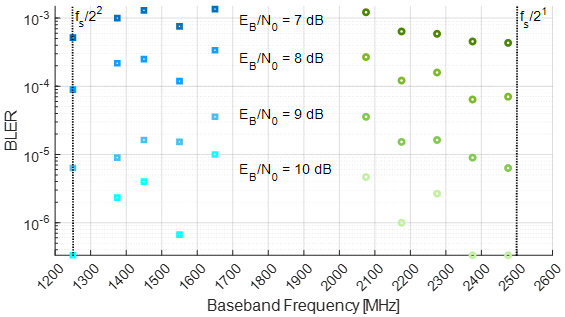}
 	\caption[BLER vs Fi]{BLER performance as a function of the interference frequency for an ICI of 6~dB}
 	\label{fig:bler-vs-bbf}
 \end{figure}

Figure~\ref{fig:bler-vs-bbf} shows the end-to-end autoencoder capability to reject an ICI of 6dB for the analyzed baseband frequency range. The figure shows the achieved BLER under the combined effect of interference and channel noise for the $\mathit{E_B/N_0}$ range where the baseline model is expected to have a $\mathit{BLER<10^{-3}}$. 
 
As the interference central frequency moves away from the vicinity of the sub-multiples of $\mathit{f_s}$, the model's capability to mitigate the impact of ICI on the BLER decreases. The average absolute gradient of the BLER per MHz deviated from  $\mathit{f_s}$ sub-multiples is $\mathit{2.36 \cdot  10^{-7}}$ for the $\mathit{E_B/N_0}$ range between 7~dB to 9~dB. The average R2 of the independent fitting functions for estimating the gradient is approximately 0.89.

\subsection{ICI rejection capability as a function of Frequency}
\begin{figure}[h]
	\centering
	\includegraphics[width=0.8\linewidth]{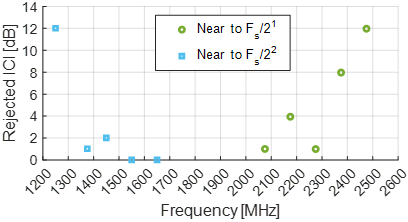}
	\caption[ICI Rejection vs Frequency]{ICI Rejection as a function of frequency}
	\label{fig:icireject-vs-freq}
\end{figure}

Figure~\ref{fig:icireject-vs-freq} shows the maximum ICI that the autoencoder can compensate to achieve a $BLER < 10^{-4}$ for a channel baseline noise equivalent to an $E_B/N_0 = 8~dB$. The end-to-end autoencoder is capable of significantly mitigating the impact of 12~dB ICI when the interference is centered near the first ($\mathit{f_s/2^1}$) and second ($\mathit{f_s/2^2}$) sub-multiples of $\mathit{f_s}$. Despite this high level of ICI, the targeted $\mathit{BLER < 10^{-4}}$ for a baseline $\mathit{E_B/N_0}$ of 8~dB is maintained. For an interference frequency in the vicinity of the first sub-multiple (-125~MHz), the auto-encoder can still reject approximately 8~dB of ICI. This drops further to 4~dB to 1~dB for larger frequency deviations (cf. green markers in Figure~\ref{fig:icireject-vs-freq}). For an interference frequency centered beyond 100 MHz from the second sub-multiple, the autoencoder can mitigate the impact of ICI up to 2~dB. However, an interference centered beyond 300 MHz from the second submultiple cannot be effectively mitigated to keep the targeted performance (cf. blue markers in Figure~\ref{fig:icireject-vs-freq}).

This is consistent with the power distribution findings from Figure~\ref{fig:interferencepatternn1} and Figure~\ref{fig:interferencepatternn2}. For the interference center frequencies that have a Walsh power distribution concentrated on a few coefficients, the end-to-end autoencoder adapts symbols to Walsh channel-uses mapping in a way that minimizes the error rate. Therefore, the encoding function tends to de-emphasize Walsh coefficients that are more susceptible to interference, as determined by the interference distribution in that domain. When the power is more uniformly distributed, the autoencoder cannot find a symbol-to-Walsh encoding function that minimizes the effect of ICI.

\section{Conclusion}
Our paper presented a Walsh-domain end-to-end wireless autoencoder capable of rejecting partial-in-band interference from coexisting 5G transmissions in UWB downlinks. Our research demonstrated that the interference has an asymmetrical power distribution pattern in the Walsh domain. For instance, near sub-multiples of the system sampling frequency, the power concentrates mostly on a few Walsh branches. We leverage this feature for dynamic spectrum assignment of the UWB signal to make 5G interference fall near these sub-multiples. An End-to-End autoencoder learns an interference-aware symbol encoding function that minimizes BLER. Simulation results confirmed that the proposed approach can reject up to 12~dB of ICI for the same baseline channel noise level. For a maximum ICI of 6~dB (typical on dense mobile deployments), the Walsh-based wireless autoencoder achieves near identical BLER performance to the reference baseline model without interference (i.e., for interference located near the sub-multiples of the sampling frequency).  These findings highlight the potential of Walsh-domain signal processing and deep learning integration for dynamic spectrum access, enabling more aggressive reuse of spectrum and coexistence with legacy technologies towards beyond 6G ultrawideband networks.

\section*{Acknowledgment}
This paper was supported by the HERMES project, which has received funding from the European Union’s Horizon 2020 research and innovation program under grant agreement No 964246. Publication reflects only the author's view; the Commission is not responsible for any use that may be made of the information it contains. The work of Rodney Martinez Alonso is supported by the Research Foundation–Flanders (FWO) under Grant 1211926N.



%\bibliographystyle{IEEEtran}

\bibliographystyle{IEEEtran}
\bibliography{mybibfile}

\end{document}